\documentclass{article}

\usepackage[utf8]{inputenc} 
\usepackage[T1]{fontenc}    
\usepackage{PRIMEarxiv}
\usepackage{fancyhdr}       
\usepackage{amssymb}
\usepackage{amsfonts}       
\usepackage{amsmath}
\usepackage{amsthm}
\usepackage{amstext}
\usepackage{nicefrac}       
\usepackage{microtype}      
\usepackage[square,numbers]{natbib}
\usepackage{color}
\usepackage{hyperref}       
\usepackage{url}            
\usepackage{booktabs}       
\usepackage{multirow}
\usepackage{subcaption}
\usepackage{lipsum}
\usepackage{graphicx}       
\usepackage{multirow}
\usepackage{booktabs}
\begin{document}
  
\title{A Benchmarking Framework for Network Classification Methods}

\author{
  Joao V. Merenda, Gonzalo Travieso, Odemir M. Bruno \\
 Artificial Intelligence and Complex System Lab \\ São Carlos Institute of Physics \\ University of São Paulo\\
  \texttt{joao.merenda@usp.br, gonzalo@ifsc.usp.br, bruno@ifsc.usp.br}}
  
\maketitle

\begin{abstract}

Network classification plays a crucial role in the study of complex systems, impacting fields like biology, sociology, and computer science. In this research, we present an innovative benchmark dataset made up of synthetic networks that are categorized into various classes and subclasses. This dataset is specifically crafted to test the effectiveness and resilience of different network classification methods. To put these methods to the test, we also introduce various types and levels of structural noise. We evaluate five feature extraction techniques: traditional structural measures, Life-Like Network Automata (LLNA), Graph2Vec, Deterministic Tourist Walk (DTW), and its improved version, the Deterministic Tourist Walk with Bifurcation (DTWB). Our experimental results reveal that DTWB surpasses the other methods in classifying both classes and subclasses, even when faced with significant noise. LLNA and DTW also perform well, while Graph2Vec lands somewhere in the middle in terms of accuracy. Interestingly, topological measures, despite their simplicity and common usage, consistently show the weakest classification performance. These findings underscore the necessity of robust feature extraction techniques for effective network classification, particularly in noisy conditions.

\end{abstract}

\section{Introduction}

Complex networks have emerged as a powerful way to understand the structure and dynamics of interconnected systems. Ranging from theoretical physics and mathematics to real-world applications in biology, sociology, and technology, studying complex networks offers valuable insights into how different elements interact within these intricate systems. Many natural and artificial systems, like social interactions, neural networks, and transportation systems, can be effectively modeled as networks, where nodes represent entities and edges illustrate their relationships \cite{yang2013networks}.

In theoretical science, complex network analysis contributes to the understanding of fundamental principles governing large-scale connectivity, resilience, and emergence of collective behavior. The study of network topology, centrality measures, and community structures has provided key insights on many types of systems, such as ecosystems, financial markets, and epidemiological models \cite{iribarren2007information, li2013modeling, myers2012information}. On the applied side, network science has proven essential for designing efficient communication infrastructures, optimizing transportation logistics, and improving recommendation systems, among other fields.

One of the main challenges in network analysis is extracting useful information from large, often noisy datasets. This is where pattern recognition techniques become essential. By utilizing feature extraction methods, machine learning models can uncover underlying patterns, detect anomalies, and classify networks based on their characteristics. The intersection of complex networks and pattern recognition has led to significant progress in areas such as network-based classification, link prediction, and graph representation learning.

In this study, we evaluate the performance of various feature extraction methods on a newly developed benchmark network dataset. By systematically comparing these methods, we aim to contribute to a deeper understanding of how different techniques capture relevant patterns in network structures. Our findings provide valuable insights for both theoretical and applied research in network science and pattern recognition.

\subsection{Background}

The study of complex networks has significantly evolved since the introduction of the Erdos-Rényi model in the 1960s, which led the foundation for understanding random graphs \cite{erd6s1960evolution}. This model posited that networks are formed by connecting nodes randomly, leading to a Poisson degree distribution. While insightful, it became evident that many real-world networks exhibit properties not captured by this model.

In 1999, László Barabási and Réka Albert unveiled the concept of scale-free networks, which are marked by a power-law degree distribution. Their findings revealed that networks, including the World Wide Web and biological systems, contain hubs with a far greater number of connections, contradicting the randomness that previous models had assumed. This idea of preferential attachment gave us a better understanding of the inherent diversity in complex systems \cite{barabasi1999emergence}.

The early 2000s marked a significant exploration into small-world networks, led by researchers Duncan Watts and Steven Strogatz. These networks are notable for their high clustering coefficients and short path lengths, much like social networks where people are interconnected through mutual friends. This model offered a deeper understanding of how information flows efficiently in both social and biological systems \cite{watts1998collective}.

As the field progressed, researchers began integrating pattern recognition techniques with complex network analysis. This interdisciplinary approach enabled the extraction of meaningful patterns from intricate datasets, enhancing tasks such as classification, anomaly detection, and prediction within interconnected systems. For instance, the application of machine learning algorithms to network data has facilitated the identification of community structures and the prediction of missing links, thereby enriching our understanding of the underlying processes governing complex networks.

This research expands on a well-established history by assessing different feature extraction methods against a newly developed benchmark network dataset. By conducting a systematic comparison, it strives to enhance our understanding of how various techniques detect relevant patterns in network structures, thereby aiding both theoretical progress and practical applications in the fields of complex networks and pattern recognition.

\section{Feature extraction methods}
\label{feat_methods}

Over the years, numerous feature extraction methods have emerged. The most traditional and straightforward approach involves computing topological measures, such as mean degree, global clustering coefficient, average shortest path length, and various centrality metrics, and using them as a feature vector. More recently, advanced methods have been developed, including Deterministic Tourist Walk (DTW), Life-Like Network Automata (LLNA), and Graph Neural Networks (GNNs), which offer more powerful representations of complex network structures.

In this section, we present the feature extraction methods evaluated in this study.

\subsection{Structural measures}

In this first method, we extract structural measures from the networks to generate a feature vector. The measures used include: average degree, hierarchical degree (levels 2 and 3), global clustering coefficient, and average shortest path length. This method is the most traditional approach for pattern recognition in networks and serves as a gold standard for benchmarking other pattern recognition methods. 

\subsection{Life-Like Network Automata}

The Life-Like Network Automata (LLNA) method is a network-based extension of cellular automata designed for pattern recognition and classification tasks. In LLNA, a network serves as the underlying structure where each vertex represents a cell with a binary state (alive or dead). The automaton evolves over discrete time steps, generating a time-evolution pattern (TEP) that encodes the system's dynamic behavior. These TEPs capture the temporal evolution of states across the network, providing a rich source of information about its structural and functional properties. By analyzing these patterns, LLNA enables the classification and characterization of different networks, making it a powerful tool for network science applications \cite{miranda2016exploring}.

The evolution of each cell’s state in LLNA is governed by a transition function that depends on its neighborhood density, which is the proportion of alive neighbors relative to its total degree as shown in equation \ref{LLNA1}. 

\begin{equation}
    \label{LLNA1}
    \sigma (v_i,t) = \frac{1}{k_i}\sum_{j=1}^N A_{ij}s(v_j,t)
\end{equation}
where, $v_i$ is the $i-th$ vertex, $k_i$ is the $i-th$ vertex degree, $A_{ij}$ is the adjacency matrix, and $s(v_j,t)$ represents the state of $j-th$ vertex at time $t$

The transition function follows Life-Like rules, denoted as $Bx-Sy$ in the form $BX_0X_1\cdots X_x-SY_0Y_1\cdots Y_y$, where $Bx$ defines the conditions under which a dead cell becomes alive (birth), and $Sy$ specifies the conditions for a living cell to remain alive (survival). 

\begin{equation}
    \label{LLNA2}
    s(c_i, t + 1) =
\begin{cases} 
1, & \text{if } s(c_i, t) = 0 \text{ and } \frac{X_x}{r} \leq \sigma(c_i, t) < \frac{X_x + 1}{r} \Rightarrow \text{born (B)} \\
1, & \text{if } s(c_i, t) = 1 \text{ and } \frac{Y_y}{r} \leq \sigma(c_i, t) < \frac{Y_y + 1}{r} \Rightarrow \text{survive (S)} \\
0, & \text{otherwise}.
\end{cases}
\end{equation}
in which, $r$ is the neighborhood size. 

 These rules, applied iteratively to all cells in the network, create a complex temporal evolution pattern (TEP), revealing intrinsic properties of the network's structure and dynamics. To further refine network characterization, the State Density Time-Evolution Pattern (SDTEP) extends the traditional TEP framework by incorporating both node states and their local density evolution over time \cite{10677825}. This enriched representation generates feature vectors constructed from SDTEP histograms, capturing essential network dynamics for classification tasks.

\subsection{Deterministic Tourist Walk}

The deterministic tourist walk (DTW) method has emerged as a significant approach for texture analysis in computer vision \cite{campiteli2006deterministic} and has recently been employed in network classification tasks, demonstrating high performance \cite{gonccalves2012complex,merenda2023using}. Unlike random walks, DTW follows a deterministic movement rule based on a specific network property. In this study, we utilize the degree difference between a node and its neighbors as the guiding criterion for movement, where the tourist seeks to either minimize or maximize this difference, following the rule $r= \text{min}$ or $r=\text{max}$, respectively.

DTW is a partially self-avoiding walk, as the tourist possesses a finite memory with size $\mu$ that prevents revisiting nodes already stored within it. The tourist's trajectory is divided into two phases: the transient phase, with size $t$, where the tourist explores the network, and the attractor phase, with size $p$, where the tourist becomes trapped in a loop, terminating the walk. Therefore, the tourist's trajectory length is $\ell = t + p$.

The method generates a distinctive signature for each network by constructing a trajectory histogram based on the lengths of the transient and attractor phases as shown in equation \ref{DTW1}, 

\begin{equation}
\label{DTW1}
    h(\ell) = \sum_{b=0}^{\ell - 1}S(b,\ell - b)
\end{equation}
where, $S(t,p)$ is a counter function defined by: 

\begin{equation}
    \label{DTW2}
    S(t,p) = \frac{1}{N}\sum_{i=1}^N
    \begin{cases}
        1 \text{, if } t_i = t \text{, and } p_i = p \\
        0 \text{, Otherwise}
    \end{cases}
\end{equation}
in which, $N$ is the number of trajectories. 

A signature for each network is created by constructing a vector (Equation \ref{DTW3}) for multiple trajectory lengths, ranging from $\mu + 1$ to $\mu + m$, where 
$m$ is an arbitrary number.

\begin{equation}
    \label{DTW3}
    \phi_{\mu}^{rule} = [h(\ell = \mu + 1),h(\ell = \mu+2),\cdots , h(\ell = \mu + m)]
\end{equation}

It has been proven that concatenating multiple memory sizes and various rules (the rule of minimum and the rule of maximun) enhances the performance of DTW \cite{backes2010texture}, leading to the construction of the following signature vector:

\begin{equation}
    \label{DTW4}
    \psi_{[\mu_1,\mu_2,\cdots \mu_n]}^{[min,max]} = [\phi_{\mu_1}^{min}, \phi_{\mu_2}^{min},\cdots , \phi_{\mu_n}^{min}, \phi_{\mu_1}^{max}, \phi_{\mu_2}^{max},\cdots , \phi_{\mu_n}^{max}]
\end{equation}

\subsection{Deterministic Tourist Walk with Bifurcation}

The deterministic tourist walk with bifurcation (DTWB) is a variation of the traditional deterministic tourist walk (DTW) designed to overcome one of its primary limitations. In the conventional DTW method, a tourist automaton with memory size $\mu$ and walking rule 
$r$ may encounter nodes where multiple neighbors satisfy the given rule. In such cases, the standard DTW randomly selects one of these neighbors, disregarding the others, which can result in information loss. The DTWB method addresses this issue by generating clones of the tourist, allowing each clone to traverse a different valid path. This approach enables a more comprehensive exploration of the network, covering additional regions and extracting richer structural information. The signature vector produced by DTWB consists of various statistical measures derived from the trajectories of these walkers \cite{merenda2025pattern}. 

The two most common walking rules for this automaton are: (i) rule $r = \text{BR1}$, in which the tourist selects neighboring nodes with the same degree as the current node; and (ii) rule $r = \text{BR2}$, in which the tourist selects neighbors with degrees different from that of the current node.

\subsection{Graph2vec model}

Graph2Vec is a representation learning technique designed to generate fixed-size embeddings for entire graphs, analogous to how Word2Vec creates vector representations for words \cite{narayanan2017graph2vec}. Unlike traditional Graph Neural Networks (GNNs), which focus on node-level embeddings by aggregating local neighborhood information, Graph2Vec learns graph-level representations by capturing structural and feature-based similarities among graphs. The method employs the Weisfeiler-Lehman (WL) subtree kernel \cite{shervashidze2011weisfeiler} to hierarchically decompose graphs into substructures, which serve as atomic units for embedding learning.

The WL algorithm iteratively refines node labels by aggregating and hashing neighborhood information, producing a sequence of increasingly refined graph representations. At each iteration $t$, nodes are assigned updated labels based on their own current label and the multiset of labels from their immediate neighbors. This process generates a hierarchy of substructures (e.g., rooted subtrees of depth $t$), encoding local-to-global structural patterns.

The core idea behind Graph2Vec is to treat entire graphs as documents and their substructures (such as node neighborhoods) as words, leveraging a skip-gram-like approach to train embeddings. The model learns to encode graphs into a continuous vector space where structurally similar graphs are placed closer together. This is achieved by optimizing an unsupervised objective that maximizes the similarity between a graph’s embedding and the representations of its substructures while minimizing similarity to unrelated graphs.

\section{Experimental setup}

\subsection{Dataset}

To evaluate the performance of the feature extraction methods presented in Section \ref{feat_methods} in generating meaningful signatures for network classification, we constructed a novel network dataset designed as a benchmark for testing various pattern recognition methods in network analysis. This dataset comprises four classes of synthetic networks, each further divided into distinct subclasses. The methods applied in this study must accurately classify the networks at both the class and subclass levels. The dataset consists of 4,200 networks distributed across the four classes. The details of these classes and subclasses are provided below. The dataset is available on http://scg.ifsc.usp.br/networksbenchmarks.

\begin{enumerate}
    \item \textbf{Random networks (R):} This class comprises 1,000 random networks belonging to two subclasses: \textit{Erdős–Rényi (ER)} and \textit{Gilbert–Erdős–Rényi (GER)}. The networks have sizes ranging from $N=100$ to $N=1,000$ nodes, with an average degree between $k=4$ and $k=22$. In the \textit{Erdős–Rényi} model, the number of edges added is defined as $m = (Nk)/2$. In the \textit{Gilbert–Erdős–Rényi} model, the connection probability is defined as $p = k/(N-1)$.
    
    \item \textbf{Small-World networks (SW):} This class consists of 1,000 networks divided into two subclasses: \textit{Watts–Strogatz (WS)}, with 500 networks, and \textit{Newman–Watts–Strogatz (NWS)}, with 500 networks. The network sizes range from $N=100$ to $N=1,000$ nodes, and an average degree between $k=4$ and $k=22$. The rewiring probability varies between $p=0.01$, and $p=0.5$.
    
    \item \textbf{Geometric networks (G):} This class has two subclasses. The networks have sizes ranging from $100$ to $1,000$ nodes, and the average degree varies according to the network parameters $\alpha$, $\beta$, and $r$.
    \begin{enumerate}
        \item \textit{Geometric Random Graph (GRG):} Includes 500 networks with connection radius varying from $r=0.05$ to $r=0.5$.
        
        \item \textit{Waxman (Wax):} Includes 500 Waxman networks with $\alpha$ varying from $\alpha = 0.25$, to $\alpha = 0.75$, and $\beta$ ranging from $\beta = 0.25$ and $\beta= 0.75$.
    \end{enumerate}

    \item \textbf{Scale Free networks (SF):} This class comprises 1,200 networks divided into five subclasses, each with sizes ranging from $100$ to $1,000$ nodes. The subclasses are: 
    
    \begin{enumerate}
        \item \textit{Barabasi-Albert (BA):} This subclass consists of 240 networks, where the number of edges added per node, $m$, varies from $4$ to $13$.

        \item \textit{Non-linear Barabasi-Albert (NBA):} This subclass consists of 240 Barabási–Albert networks with a non-linear connection rate, where the exponent $\alpha$ ranges from $1.0$ to $3.0$, and the number of connections added per step ranges from $m=4$ to $m=16$.

        \item \textit{Holme-Kim (HK)}: This subclass consists of 240 networks, where the number of edges added per node, $m$, ranges from $4$ to $13$, and the triadic Closure parameter, $p$, ranges from $0.01$ to $1.0$.

        \item \textit{Price (P):} This subclass consists of 240 networks from the Price model, where the number of edges added per node, $m$, ranges from $4$ to $13$, and the attractiveness constant, $c$ varies from $0.1$ to $3.0$.

        \item \textit{Dorogovtsev-Mendes (DMS):} This subclass consists of 240 Dorogovtsev-Mendes networks, where the aging Rate, $a$, varies from $0.5$ to $3.0$.
    \end{enumerate}

\end{enumerate}

\textbf{Noisy dataset} To evaluate the robustness of the feature extraction methods, we introduce noise into the dataset. Four types of noise were added to generate perturbations in the networks:

\begin{enumerate}
    \item \textbf{Link Adding (LA):} Random edges are added to the graph.

    \item \textbf{Link Removing (LR):} Random edges are removed from the graph.

    \item \textbf{Link Combination (LC):} This method combines the two strategies above by randomly removing an existing edge and adding a new one. This transformation preserves the average degree.

    \item \textbf{Link Switch (LS):} Two pairs of edges, $(i,j)$ and $(k,l)$, are randomly selected, and their endpoints are swapped, resulting in the new edges $(i,l)$ and $(k,j)$.
\end{enumerate}

From each network in the dataset, four noisy versions are generated using the techniques described above. As a result, the noisy dataset consists of 16,800 networks. The amount of noise introduced is controlled by the parameter $\rho$, also referred to as the \textit{Noise rate}, which quantifies the number of modifications applied to the graph. In this work, we used ten noise rates: $10\%$, $20\%$, $30\%$, $40\%$, $50\%$, $60\%$, $70\%$, $80\%$, $90\%$, and $100\%$.

\subsection{Feature Extraction Methods Setup and Parameters}

The methodologies employed in this study can be categorized into two main groups: feature extraction methods and classification methods. The feature extraction methods generate feature vectors that represent network characteristics, while the classification methods apply machine learning techniques to categorize these feature vectors. In this section, we define the setup and parameters of the feature extraction methods introduced in Section \ref{feat_methods}.

\textbf{LLNA:} For the Life-Like Network Automata, we employ the rule B1234-S456 which has been identified as optimal for synthetic networks \cite{10677825}. The feature vector is constructed using the SDTEP variant of LLNA, which represents a histogram that integrates information about both the state of a node and the states of its surrounding neighborhood.

\textbf{DTW:} For the tourist method, we employ four distinct memory sizes, $\mu = [1, 2, 3, 4]$, and two walking rules, $r = [\text{min}, \text{max}]$. We evaluated DTW performance using one memory size and one rule at a time ($\phi_{\mu}^{r}$) and the combination of all memory sizes and rules ($\psi_{[1,2,3,4]}^{[\text{min},\text{max}]}$).

\textbf{DTWB:} For the DTWB method, we adopt a fixed memory size, $\mu = 1$, and two distinct walking rules, $[\text{BR1}, \text{BR2}]$. Previous works have shown that the performance of the DTWB method does not vary with the memory size, only with the walking rule \cite{merenda2025pattern}. Feature vector extraction is performed by applying each rule individually as well as in combination. To formulate the feature vector, we utilize the statistical measures outlined in the methodology proposed by Merenda \textit{et al.} \cite{merenda2025pattern}. 

\textbf{Graph2vec:} The Graph2Vec model was implemented using the karateclub library, with graph embeddings generated through the Weisfeiler-Lehman (WL) subtree kernel method. The model parameters included 128-dimensional embeddings and two WL iterations (wl-iterations=2) to iteratively aggregate structural features from node neighborhoods. This configuration aimed to capture hierarchical graph topology while balancing computational efficiency and representational granularity.

\subsection{Evaluation methods}

To evaluate the performance of the methods above, we use a \textit{Support Vector Machine} (SVM) as the classification method, with a \textit{Radial Basis Function} (RBF) kernel, which is well-suited for classifying non-linear data. The regularization parameter is set to $C = 1.0$, and the kernel parameter $\gamma$ is set to 'scale', which automatically adjusts based on the data distribution.

To validate the methods, we apply the Leave-One-Out Cross-Validation (LOOCV) strategy, which provides an unbiased estimate of model performance. LOOCV is a special case of k-fold cross-validation where each sample in the dataset is used as a test set exactly once, while the remaining $N-1$ samples serve as the training set. This process repeats $N$ times, ensuring that every data point is tested independently.

The evaluation process of the dataset can be divided into two steps:

\begin{enumerate}
    
\item Noiseless Dataset: To assess the performance of the feature extraction methods in a noise-free environment, we apply the Support Vector Machine (SVM) classifier using the Leave-One-Out Cross-Validation (LOOCV) strategy. This allows us to compute the mean accuracy and standard deviation of the classification task.

\item Noisy Datasets: In this step, we train the SVM model on the noiseless dataset and test it on the noisy dataset. The goal is to evaluate the robustness of the methods, assessing their ability to differentiate between noisy and noise-free networks.

\end{enumerate}

\section{Results and discussion}

According to Table \ref{table2}, the highest performance in classifying networks into the four synthetic models (Random, Scale-Free, Small-World, and Geometric) was achieved by the DTWB method. This result was obtained using a combination of two walking rules (BR1 and BR2) and a memory size of $\mu = 1$, reaching an accuracy of $99.50\%$. The DTWB method also performed best in the subclass classification task, correctly identifying both the class and subclass of a network with an accuracy of $86.95\%$. The second and third best performances were achieved by the DTW method (with combinations of walking rules and memory sizes) and the LLNA-SDTEP method, respectively. In contrast, the worst performance came from the topological measures, which reached an accuracy of only $83.37\%$.

\begin{table}[h!]
\centering
\begin{tabular}{lcc|cc}
\toprule
 & \multicolumn{2}{c|}{Class} & \multicolumn{2}{c}{Subclass} \\
\cmidrule(r){2-3} \cmidrule(l){4-5}
\textbf{Methods} & \textbf{Avg. accuracy (\%)} & \textbf{Deviation (\%)} & \textbf{Avg. accuracy (\%)} & \textbf{Deviation (\%)} \\
\midrule
$\phi (1,min)$ & 63.66 & 8.10 & 52.20 & 19.95 \\
$\phi (2,min)$ & 48.79 & 9.99 & 37.51 & 18.42 \\
$\phi (3,min)$ & 45.88 & 9.83 & 31.41 & 16.30 \\
$\phi (4,min)$ & 43.52 & 9.58 & 26.76 & 16.94 \\
$\phi (1,max)$ & 76.10 & 2.65 & 54.35 & 19.81 \\
$\phi (2,max)$ & 56.89 & 9.52 & 35.23 & 17.49 \\
$\phi (3,max)$ & 56.29 & 8.23 & 34.52 & 17.81 \\
$\phi (4,max)$ & 56.59 & 9.56 & 28.82 & 15.29 \\
$\psi (\Vec{\mu},\Vec{r})$ & 98.88 & 12.73 & 75.53 & 12.99 \\
$\phi_B (1,BR1)$ & 97.62 & 7.06 & 68.58 & 42.42 \\
$\phi_B (1,BR2)$ & 97.78 & 14.72 & 83.78 & 36.87 \\
$\psi_B (1,\Vec{r})$ & 99.50 & 7.06 & 86.95 & 33.69 \\

\bottomrule
\end{tabular}
\caption{Class and subclass performance of the Tourist methods (DTW and DTWB). The left column presents the average accuracy and corresponding standard deviation for the class classification task, while the right column shows the same metrics for subclass classification. The feature vector $\phi$ is defined in Equation \ref{DTW3}, and $\psi$ in Equation \ref{DTW4}. The subscripts $\phi_B$ and $\psi_B$ refer to the DTWB method. The rule vector $\Vec{r}$ is defined as $[\text{min}, \text{max}]$ for DTW, and $[\text{BR1}, \text{BR2}]$ for DTWB. The memory size vector $\Vec{\mu}$ is given by $[1, 2, 3, 4]$.}
\label{table1}
\end{table}

The results presented in Table \ref{table1} show that, for the noiseless dataset, the rule $r=max$ outperforms the rule $r=min$ when applied to the DTW method. Additionally, the accuracy tends to decrease as the memory size increases. However, this trend is not entirely observed in the noisy dataset (Table \ref{table3}). In both cases, noiseless and noisy, the performance improves when multiple memory sizes and walking rules are combined, both in the DTW and DTWB methods. This enhancement is consistent with findings from previous studies \cite{gonccalves2012complex,merenda2025pattern}.

Additionally, the subclass classification task exhibited a higher standard deviation, indicating greater variability and stronger data dependence. This suggests that subclass classification is a more challenging task for the evaluated methods.

\begin{table}[h!]
\centering
\begin{tabular}{lcc|cc}
\toprule
 & \multicolumn{2}{c|}{Class} & \multicolumn{2}{c}{Subclass} \\
\cmidrule(r){2-3} \cmidrule(l){4-5}
\textbf{Methods} & \textbf{Avg. accuracy (\%)} & \textbf{Deviation (\%)} & \textbf{Avg. accuracy (\%)} & \textbf{Deviation (\%)} \\
\midrule
LLNA & 98.24 & 13.16 & 76.80 & 18.94 \\
Measures & 83.37 & 17.23 & 62.51 & 18.29 \\
Graph2vec & 96.43 & 8.46 & 68.05 & 1.69 \\
$\psi_B (1,\Vec{r})$ & 99.50 & 7.06 & 86.95 & 33.69 \\
$\psi (\Vec{\mu},\Vec{r})$ & 98.88 & 12.73 & 75.53 & 12.99 \\

\bottomrule
\end{tabular}
\caption{Class and subclass performance of the presented methods. The left column presents the average accuracy and corresponding standard deviation for the class classification task, while the right column shows the same metrics for subclass classification. These metrics were obtained using the k-fold cross-validation method.}
\label{table2}
\end{table}

\begin{table}[h!]
\centering
\begin{tabular}{lcccccccccc}
\toprule
 & \multicolumn{10}{c}{Noise rate (\%)} \\
\cmidrule(l){2-11}
\textbf{Methods} & \textbf{10} & \textbf{20} & \textbf{30} & \textbf{40} & \textbf{50} & \textbf{60} & \textbf{70} & \textbf{80} & \textbf{90} & \textbf{100} \\
\midrule
$\phi (1,min)$ & 44.15 & 42.82 & 42.21 & 41.69 & 41.29 & 41.09    & 40.72  & 40.46  &  39.52  &  39.18 \\
$\phi (2,min)$ & 58.53 & 54.57 & 51.03 & 47.25 & 45.14 & 43.95 &     42.32 & 40.37 & 38.48 & 38.50\\
$\phi (3,min)$ & 53.57 & 50.11 & 45.79 & 43.35 & 41.73 & 40.21 &     38.48 & 37.46 & 35.99 & 35.57 \\
$\phi (4,min)$ & 49.41 & 45.56 & 42.11 & 40.51 & 39.47 & 38.54  & 37.66 & 36.69 & 35.54& 35.31 \\
$\phi (1,max)$ & 66.51 & 65.25 & 63.94 & 61.87 & 60.08 & 57.48 & 54.10 & 49.85 & 44.69 & 45.10 \\
$\phi (2,max)$ & 51.16 & 52.14 & 52.18 & 50.76 & 49.72 & 48.12 & 46.24 & 45.20 & 42.21 & 42.07\\
$\phi (3,max)$ & 57.90 & 58.11 & 56.82 & 54.52 & 52.68 & 49.62 & 47.05 & 44.41 & 41.16 & 42.43 \\
$\phi (4,max)$ & 53.78 & 53.14 & 50.18 & 47.08 & 45.45 & 43.55    & 42.16  &  40.98   & 38.37  &  39.99 \\
$\psi (\Vec{\mu},\Vec{r})$ & 98.35  &   97.29  &  95.88 &  94.73  &  91.44  &   88.88  &   86.14  &  85.74  &  83.18  &  83.12 \\
$\phi_B (1,BR1)$ & 97.39 & 95.73 & 93.46 & 90.77 & 87.19 & 83.92 & 80.23 & 73.73  & 66.06 & 72.19   \\
$\phi_B (1,BR2)$ & 94.54 & 92.57 & 90.17 &  88.21 & 86.02 & 84.04 & 81.25 & 77.73 & 72.60 & 71.42 \\
$\psi_B (1,\Vec{r})$ & 98.94 & 97.93 & 96.20 &  94.45 & 91.32 & 88.28 & 85.44 & 81.95 & 76.29 & 75.73 \\
LLNA & 97.03 & 96.42 & 94.66 & 93.96 & 90.75  & 87.41 & 82.17 & 77.00 & 71.23 & 66.03 \\
Graph2vec & 98.13  & 97.36 & 96.25 & 94.45 &  92.12  & 89.89 & 87.35 & 83.58 & 77.79 &  64.70 \\
Measures & 83.71 & 81.49 &  77.20 & 71.98 & 65.27 & 60.34 & 56.80 & 50.17 & 48.21  & 44.44 \\
\bottomrule
\end{tabular}
\caption{Performance of the methods in class classification under varying noise levels.}
\label{table3}
\end{table}

\begin{table}[h!]
\centering
\begin{tabular}{lcccccccccc}
\toprule
 & \multicolumn{10}{c}{Noise rate (\%)} \\
\cmidrule(l){2-11}
\textbf{Methods} & \textbf{10} & \textbf{20} & \textbf{30} & \textbf{40} & \textbf{50} & \textbf{60} & \textbf{70} & \textbf{80} & \textbf{90} & \textbf{100} \\
\midrule
$\phi (1,min)$ & 27.39 & 24.53 & 23.00 & 21.77 & 21.55 & 23.12 &    22.43 &  19.32 & 18.68 & 21.55 \\
$\phi (2,min)$ & 33.11 & 29.36 & 27.11 &  25.67 & 24.08 &     22.89 & 21.70 & 20.68 & 19.31 & 19.28\\
$\phi (3,min)$ & 29.33 & 26.00 & 24.56 & 22.57 & 21.37 & 20.57 &    19.50 & 18.51 &   16.20  &  16.01\\
$\phi (4,min)$ & 26.24 & 23.47 & 21.51 & 20.19 & 18.95 & 18.40 &    17.53 &  16.58 & 15.23 & 15.92 \\
$\phi (1,max)$ & 40.09 & 37.73 & 35.56 &  33.80 & 32.51 &    30.03 & 27.93 & 26.01 &  23.34 & 26.65 \\
$\phi (2,max)$ & 34.07 & 30.33 & 29.21 & 26.86 & 25.18 & 23.32 &    22.18 & 21.52 &  19.82  &  19.87 \\
$\phi (3,max)$ & 34.13  & 31.02 & 28.77 & 26.41 & 24.77 &   23.26  & 21.85 & 20.63  & 18.84 &  19.30 \\
$\phi (4,max)$ & 30.30  &   26.85  &  23.18  & 21.24 &   20.53 &    19.98  &   19.17  &  18.43  &  16.86 &   18.05 \\
$\psi (\Vec{\mu},\Vec{r})$ & 63.38   &  59.12  &  55.26 &  52.28   & 49.09  &   45.81  &   42.36  &  39.35  &  35.73 &   39.88 \\
$\phi_B (1,BR1)$ & 60.95 & 57.97 & 56.04 & 52.69 & 50.77 & 47.64 & 45.07 & 42.39 & 38.94 & 43.75  \\
$\phi_B (1,BR2)$ & 64.41 & 61.68 & 57.94 & 55.65 & 52.66 & 50.14  & 47.02 & 44.47 & 42.60 & 45.64  \\
$\psi_B (1,\Vec{r})$ & 67.86 & 65.75 & 63.35 & 61.42 &  59.11 & 55.89 & 52.81 & 50.00 & 46.50 & 49.78 \\
LLNA & 70.60 & 66.61 & 62.45 & 59.89 & 56.15 & 53.77 & 48.80 & 45.29 & 40.87 & 45.51 \\
Graph2vec & 67.21 & 64.52 & 61.80 & 59.83 & 58.57 & 56.68 &   54.60 & 51.80 & 48.66 &  39.52 \\
Measures & 63.29 & 61.39 & 56.50 & 53.07 & 50.15 & 46.27 &     44.28  & 41.89 & 37.14 & 36.98 \\
\bottomrule
\end{tabular}
\caption{Performance of the methods in class classification under varying noise levels.}
\label{table4}
\end{table}

\begin{figure}[htbp]
    \centering
    \includegraphics[width=1.0\linewidth]{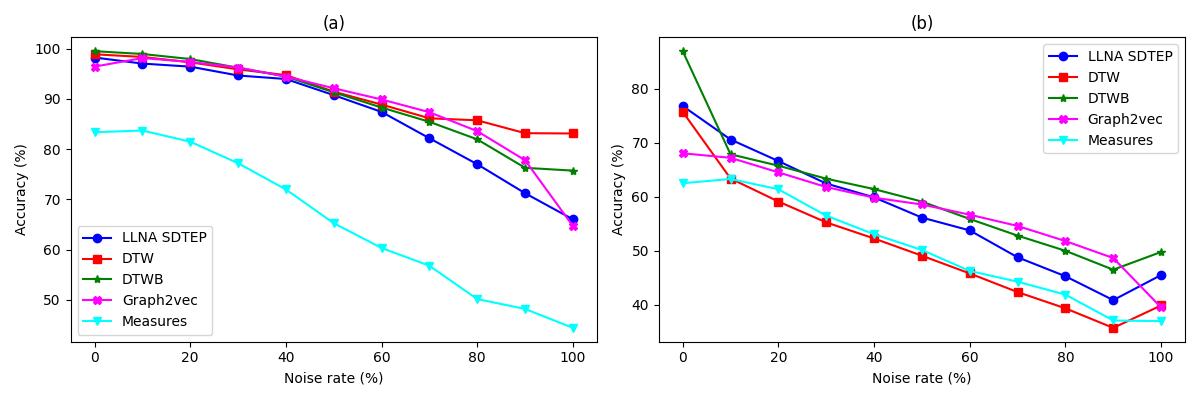}  
    \caption{Evolution of classification performance as noise is added. (a) Class classification task. (b) Subclass classification task.}
    \label{fig1}
\end{figure}

Tables \ref{table3} and \ref{table4}, along with Figure \ref{fig1}, show that performance naturally decreases as the noise rate increases. In terms of class classification, the traditional DTW method demonstrated greater robustness to noise, while the DTWB method performed better in subclass classification under noisy conditions. This suggests that DTW may be more effective at capturing general features associated with broader classes, whereas DTWB, due to its bifurcation structure, is better suited for extracting finer, more specific patterns found in subclasses by probing deeper into the network structure.

Topological measures exhibited an almost linear decline in performance with increasing noise. Although they lagged behind other methods in class classification, they achieved comparable performance in subclass classification under high noise levels.

Interestingly, a slight increase in subclass classification accuracy was observed for the DTW, DTWB, and LLNA methods between noise levels of $\rho = 90\%$ and $\rho = 100\%$. The cause of this unexpected improvement remains unclear and warrants further investigation.

\begin{figure}[htbp]
    \centering
    \includegraphics[width=1.0\linewidth]{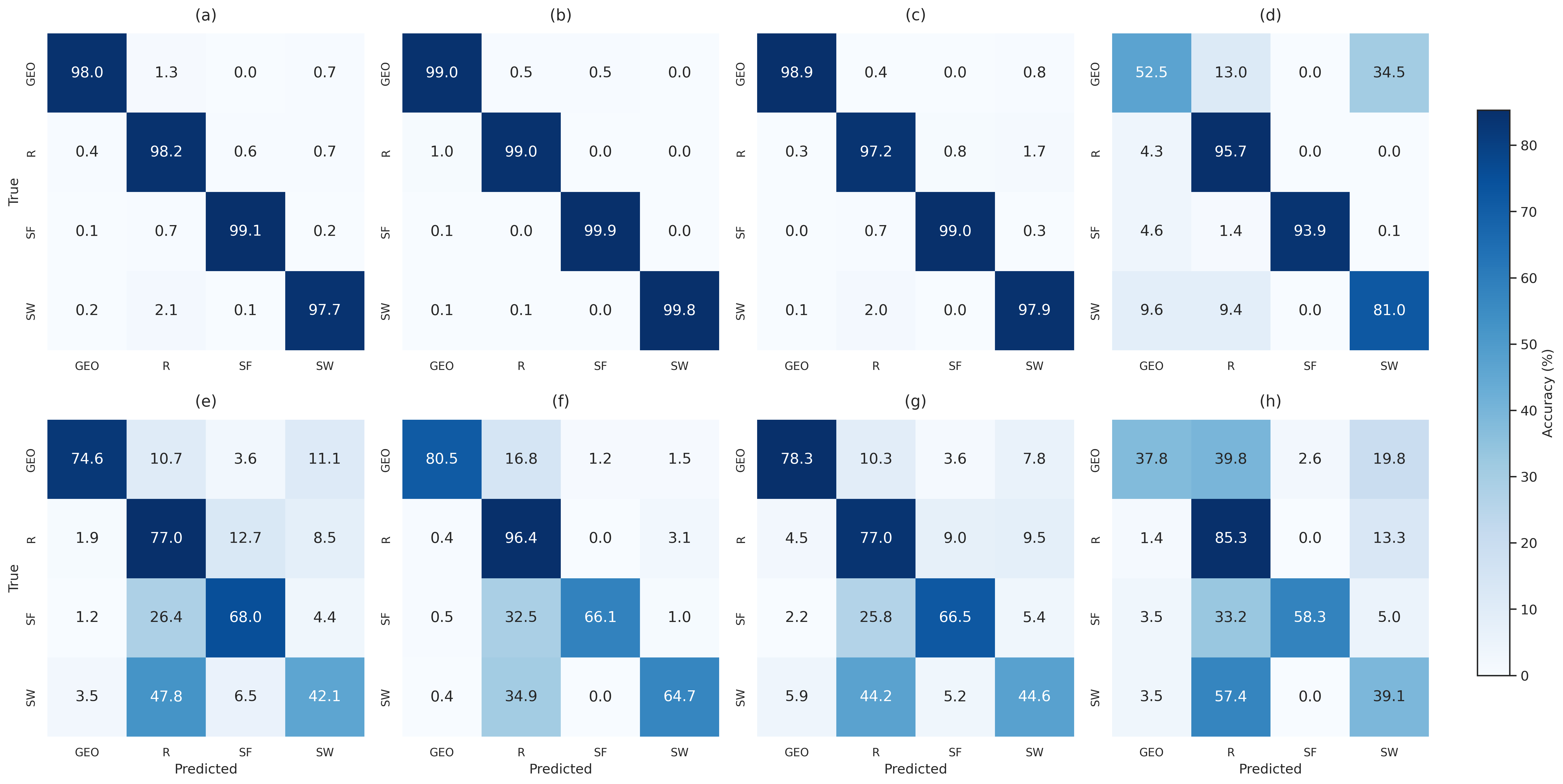}  
    \caption{Confusion matrices for the methods evaluated in this study, showing classification performance across the four synthetic network classes: Random (R), Small-World (SW), Scale-Free (SF), and Geometric (G). The top row displays the results for the noiseless dataset, while the bottom row shows the outcomes with $100\%$ noise added. Subfigures: (a) and (e) correspond to the DTW method; (b) and (f) to the DTWB method; (c) and (g) to the LLNA method; and (d) and (h) to the topological measures.}
    \label{fig2}
\end{figure}

\begin{figure}[htbp]
    \centering
    \includegraphics[width=1.0\linewidth]{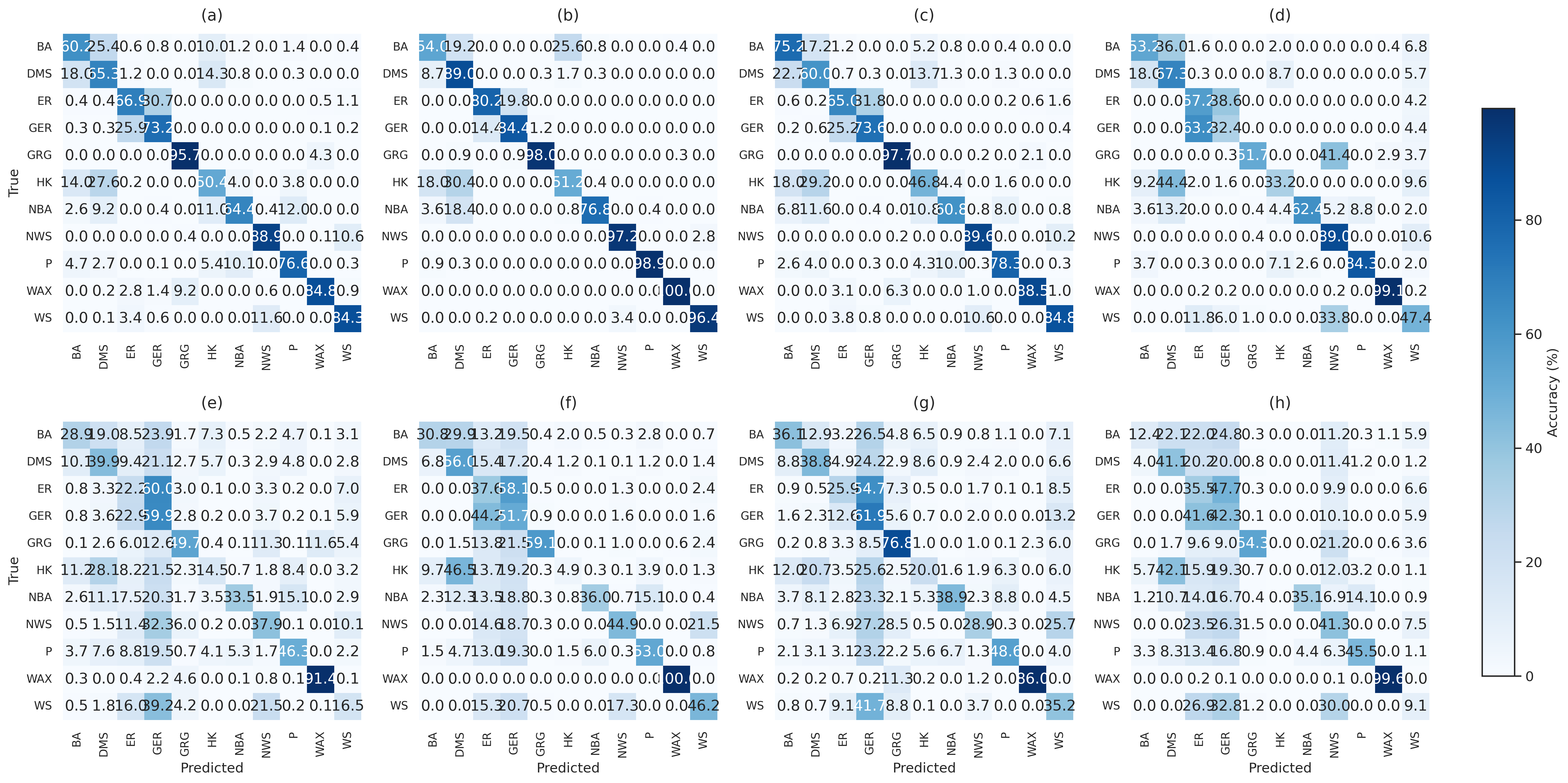}  
    \caption{Confusion matrices for the methods evaluated in this study, showing classification performance across the eleven network subclasses: Erdos-Renyi (ER), Gilbert-Erdos-Renyi (GER), Watts-Strogatz (WS), Newman-Watts-Strogatz (NWS), Geometric Random Graph (GRG), Waxman (Wax), Barabasi-Albert (BA), Non-linear Barabasi-Albert (NBA), Holme-Kim (HK), Price (P), and Dorogovtsev-Mendes (DMS). The top row displays the results for the noiseless dataset, while the bottom row shows the outcomes with $100\%$ noise added. Subfigures: (a) and (e) correspond to the DTW method; (b) and (f) to the DTWB method; (c) and (g) to the LLNA method; and (d) and (h) to the topological measures.}
    \label{fig3}
\end{figure}

Figures \ref{fig2} and \ref{fig3} display the confusion matrices for class and subclass classification, respectively. The first row of each figure shows the performance of the methods before noise was added. In this noiseless scenario, the majority of predictions lie along the main diagonal, indicating high accuracy in correctly classifying the classes and subclasses. In contrast, the second row illustrates the effect of noise: predictions become more dispersed, with increased misclassifications across incorrect classes and subclasses, reflecting the performance degradation caused by noise.

\section{Conclusion}

Developing methods for network classification is a fundamental task with numerous real-world applications. Equally important is the creation of synthetic network datasets that emulate key properties of real networks in order to challenge and evaluate these classification methods. In this study, we introduced a novel synthetic network dataset, structured into classes and subclasses, specifically designed for benchmarking network classification approaches. Additionally, we incorporated varying levels of noise to further assess the robustness of the methods.

Among the evaluated techniques, the DTW, DTWB, and LLNA methods demonstrated high accuracy in both class and subclass classification tasks, along with strong resilience to noise. In contrast, the use of topological measures, while simple and widely adopted, yielded the lowest performance. The Graph2Vec method, based on neural networks, achieved intermediate results, highlighting the trade-off between complexity and effectiveness in different approaches.

\section{Acknowledgments}
O. M. B. acknowledges support from FAPESP (grant \#21/08325-2) and CNPq (grant \#305610/2022-8). 
J. M. acknowledges support from CAPES.


\begin{thebibliography}{10}

\bibitem{yang2013networks}
Song Yang.
\newblock Networks: An introduction by mej newman: Oxford, uk: Oxford university press. 720 pp., 2013.

\bibitem{iribarren2007information}
Jos{\'e}~Luis Iribarren and Esteban Moro.
\newblock Information diffusion epidemics in social networks.
\newblock {\em arXiv preprint arXiv:0706.0641}, 2007.

\bibitem{li2013modeling}
Dong Li, Zhiming Xu, Yishu Luo, Sheng Li, Anika Gupta, Katia Sycara, Shengmei Luo, Lei Hu, and Hong Chen.
\newblock Modeling information diffusion over social networks for temporal dynamic prediction.
\newblock In {\em Proceedings of the 22nd ACM international conference on Information \& Knowledge Management}, pages 1477--1480, 2013.

\bibitem{myers2012information}
Seth~A Myers, Chenguang Zhu, and Jure Leskovec.
\newblock Information diffusion and external influence in networks.
\newblock In {\em Proceedings of the 18th ACM SIGKDD international conference on Knowledge discovery and data mining}, pages 33--41, 2012.

\bibitem{erd6s1960evolution}
Paul Erd6s and Alfr{\'e}d R{\'e}nyi.
\newblock On the evolution of random graphs.
\newblock {\em Publ. Math. Inst. Hungar. Acad. Sci}, 5:17--61, 1960.

\bibitem{barabasi1999emergence}
Albert-L{\'a}szl{\'o} Barab{\'a}si and R{\'e}ka Albert.
\newblock Emergence of scaling in random networks.
\newblock {\em science}, 286(5439):509--512, 1999.

\bibitem{watts1998collective}
Duncan~J Watts and Steven~H Strogatz.
\newblock Collective dynamics of ‘small-world’networks.
\newblock {\em nature}, 393(6684):440--442, 1998.

\bibitem{miranda2016exploring}
Gisele Helena~Barboni Miranda, Jeaneth Machicao, and Odemir~Martinez Bruno.
\newblock Exploring spatio-temporal dynamics of cellular automata for pattern recognition in networks.
\newblock {\em Scientific Reports}, 6(1):37329, 2016.

\bibitem{10677825}
Kallil M.~C. Zielinski, Leonardo Scabini, Lucas~C. Ribas, and Odemir~M. Bruno.
\newblock Exploring neighborhood variancy for rule search optimization in life-like network automata.
\newblock In {\em 2024 14th International Conference on Pattern Recognition Systems (ICPRS)}, pages 1--7, 2024.

\bibitem{campiteli2006deterministic}
M{\^o}nica~G Campiteli, Pablo~D Batista, Osame Kinouchi, and Alexandre~S Martinez.
\newblock Deterministic walks as an algorithm of pattern recognition.
\newblock {\em Physical Review E—Statistical, Nonlinear, and Soft Matter Physics}, 74(2):026703, 2006.

\bibitem{gonccalves2012complex}
Wesley~Nunes Gon{\c{c}}alves, Alexandre~Souto Martinez, and Odemir~Martinez Bruno.
\newblock Complex network classification using partially self-avoiding deterministic walks.
\newblock {\em Chaos: An Interdisciplinary Journal of Nonlinear Science}, 22(3), 2012.

\bibitem{merenda2023using}
Jo{\~a}o~VBS Merenda and Odemir~M Bruno.
\newblock Using deterministic self-avoiding walks as a small-world metric on watts--strogatz networks.
\newblock {\em Physica A: Statistical Mechanics and its Applications}, 621:128713, 2023.

\bibitem{backes2010texture}
Andr{\'e}~Ricardo Backes, Wesley~Nunes Gon{\c{c}}alves, Alexandre~Souto Martinez, and Odemir~Martinez Bruno.
\newblock Texture analysis and classification using deterministic tourist walk.
\newblock {\em Pattern Recognition}, 43(3):685--694, 2010.

\bibitem{merenda2025pattern}
Joao~V Merenda, Gonzalo Travieso, and Odemir~M Bruno.
\newblock Pattern recognition on networks using bifurcated deterministic self-avoiding walks.
\newblock {\em Chaos, Solitons \& Fractals}, 194:116100, 2025.

\bibitem{narayanan2017graph2vec}
Annamalai Narayanan, Mahinthan Chandramohan, Rajasekar Venkatesan, Lihui Chen, Yang Liu, and Shantanu Jaiswal.
\newblock graph2vec: Learning distributed representations of graphs.
\newblock {\em arXiv preprint arXiv:1707.05005}, 2017.

\bibitem{shervashidze2011weisfeiler}
Nino Shervashidze, Pascal Schweitzer, Erik~Jan Van~Leeuwen, Kurt Mehlhorn, and Karsten~M Borgwardt.
\newblock Weisfeiler-lehman graph kernels.
\newblock {\em Journal of Machine Learning Research}, 12(9), 2011.

\end{thebibliography}

\end{document}